\documentstyle[psfig]{aipproc}

\begin{document}
\title{BATSE GRB Location Errors}

\author{
M. S. Briggs$^*$,
G. N. Pendleton$^*$,  
J. J. Brainerd$^*$,   \\
V. Connaughton$^\dagger$, 
R. M. Kippen$^*$,
C. Meegan$^\ddagger$,
K. Hurley$^\star$}
\address{
$^*$UAH,
$^\dagger$NRC/MSFC,
$^\ddagger$MSFC,
$^\star$UCB}

%\lefthead{LEFT head}
%\righthead{RIGHT head}
\maketitle

\begin{abstract}
We characterize the error distribution of BATSE GRB locations
by modeling the distribution of separations between BATSE locations
and IPN annuli.
We determine error model parameters by
maximizing likelihood and rank the models
by their Bayesian odds ratios.
The best models have several systematic error terms.
The simplest good model has a 1.9 degree systematic error with probability
73\% and 5.4 degrees with probability 27\%.
A more complex model adds a dependence on the datatype used to derive
the location.

\end{abstract}

\section*{Analysis Method}

The purpose of this paper is to develop
better models of the  error distribution of burst locations in
the 3B [5] and 4B [6]
catalogs, all of which were obtained
with the same algorithm.
A more accurate error distribution will aid counterpart searches
and improve the accuracy of limits on burst repetition and clustering.

Measuring the actual location errors requires a comparison dataset
of sufficient accuracy and size---the most suitable set is the
locations of the Interplanetary Network (IPN) [4].
The advantages of this dataset are its  large size
and small errors.
The disadvantages are that 
most of the error boxes are single annuli 
and that the sample is biased towards bright GRBs.

After removing 11 wide annuli ($3\sigma$ error $> 0.8^\circ$),
the IPN supplement [4] to the 4B catalog consists of 442 GRBs.
Of these, 30 have intersecting annuli that yield essentially
`point' locations and thus provide measurements of the angles $\gamma$ between
the true locations and the BATSE locations.
For the remaining events with only single annuli we can 
determine only the closest approach
angles $\rho$ between the annuli and the corresponding
BATSE locations.  

Our method of analyzing this set of $\gamma$ and $\rho$ measurements
is similar to that of 
Graziani \& Lamb [3].
We assume various models for the systematic error $\sigma_{\rm sys}$ and
obtain the total error $\sigma_{\rm tot}$ as the rms sum of 
the statistical error $\sigma_{\rm stat}$ listed in the catalogs and
$\sigma_{\rm sys}$.
For the probability distribution $P(\gamma)$ we assume the Fisher 
distribution, which is 
generally considered to be the equivalent of the Gaussian distribution
on the sphere [2].
For those GRBs with single annuli for which we know
$\rho$ but not $\gamma$, we derive the exact distribution $P(\rho)$
from $P(\gamma)$ [1].

The model parameters are obtained  by maximizing the likelihood,
which is the product of the probability, according to the model,
of the observations:
        \begin{equation}
            L = \prod_{i} P(\gamma_i)  \prod_{j} P(\rho_j),
        \end{equation}

Once the likelihood is optimized for each model, the models are
compared by their Bayesian odds ratio [8]: 
        \begin{equation}
              O_{\rm B/A} =  \frac {P({\rm B})} {P({\rm A})}
      =  \frac {P_{\rm prior}({\rm B}) \times L({\rm B}) \times {\rm Occam(B)}}
               {P_{\rm prior}({\rm A}) \times L({\rm A}) \times {\rm Occam(A)}}
        \end{equation}
We set the first factor, representing our prior preferences for the models,
to unity.
The second factor, the likelihood ratio,
indicates how strongly the data favor one model over
another.    The ``Occam's factor''
penalizes a model to the extent that the
better fit is obtained by additional parameters.
For the data and models presented below, the factor per parameter by which
a model is penalized ranges from  about  5 to 20 so that the 
dominant factors in distinguishing between the models are the likelihood ratios.

\section*{Results}

We first try the minimal model (Model~0), 
which has
$\sigma_{\rm sys} = 1.6^\circ$ for all GRBs.
The value $1.6^\circ$ was obtained
using 50 BATSE locations and the corresponding
WATCH, COMPTEL and IPN locations as the rms difference between the actual
separations $\gamma$
and the errors estimated from the errors of the other instrument
and the BATSE statistical errors [5,7].
Figure~1 shows that the agreement of the
minimal model with the data is very poor.
Model~1 generalizes Model~0 by making $\sigma_{\rm sys}$ a
free parameter with the same value for all GRBs.
While a vast improvement over Model~0 (see Table~1), the agreement between
Model~1 and the data is still quite poor (histogram not shown).

\begin{figure}[tb!]
\mbox{
\hspace{2.5mm}	
\psfig{figure=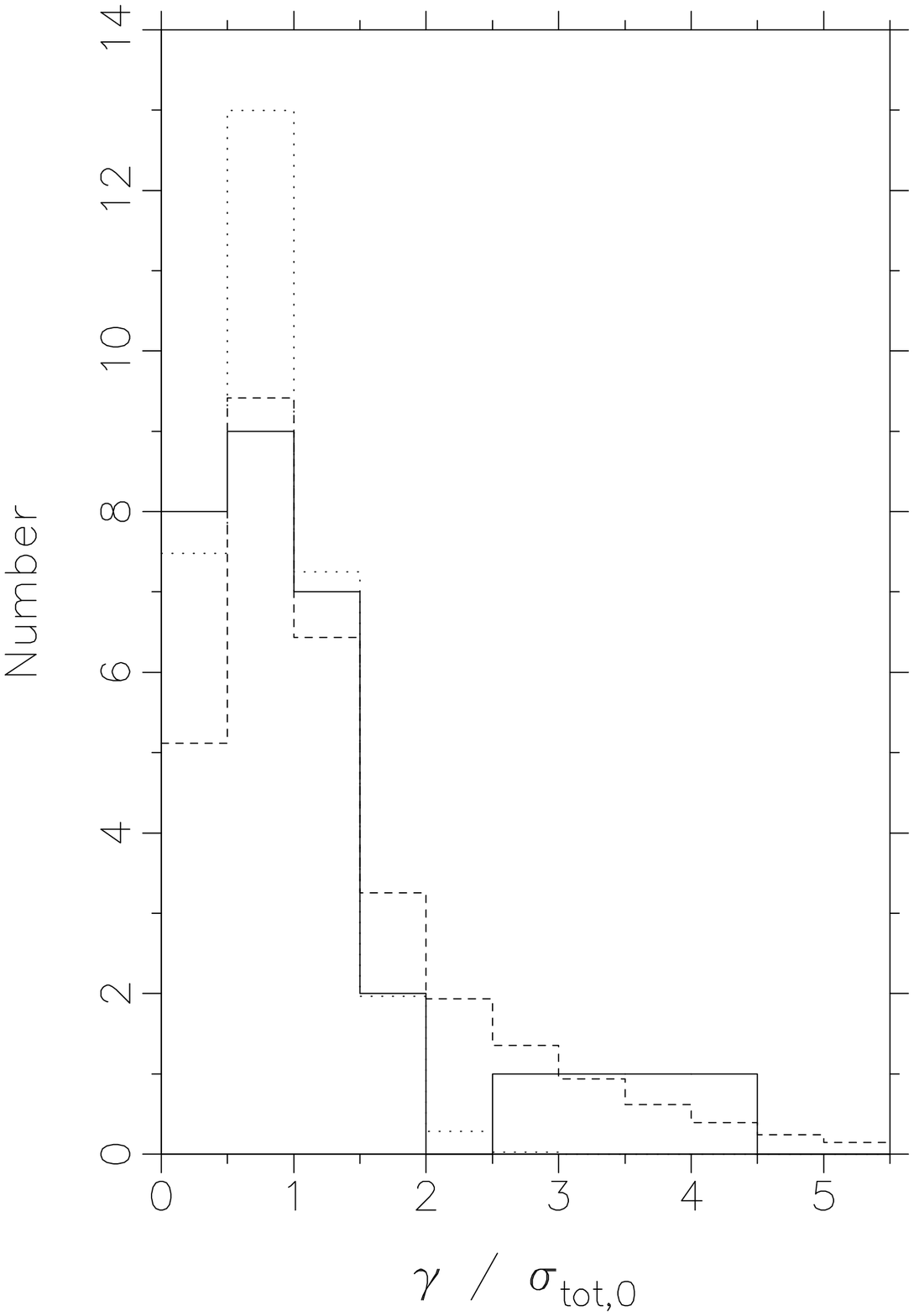,width=61mm,%
bbllx=30bp,bblly=40bp,bburx=495bp,bbury=690bp,clip=}
\hspace{4mm}
\psfig{figure=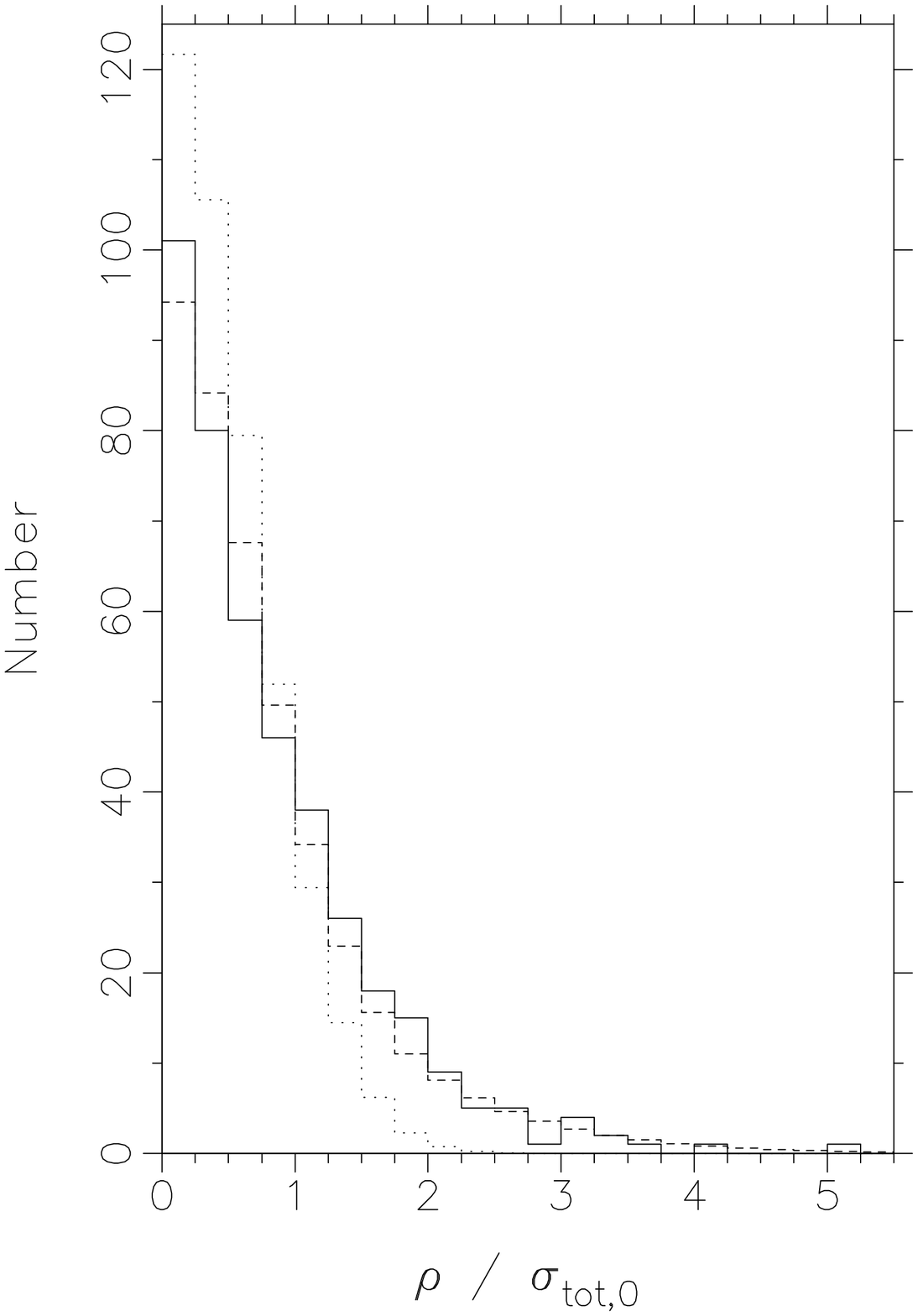,width=61mm,%
bbllx=30bp,bblly=40bp,bburx=495bp,bbury=690bp,clip=}
\hspace{\fill}
}
\vspace{1.0mm}
\caption{Histograms of the data and models.
Left: 30 events for which two annuli are available.
Right: 412 events for which only single annuli are available.
The data and both models are binned in units of $\sigma_{{\rm tot},0}$,
as calculated with the minimal model (Model~0)
of $\sigma_{\rm sys} = 1.6^\circ$.
The solid lines show the data,
the dotted lines the minimal model, and the dashed lines show Model~12.
The histogram for Model~2 (not shown) is very similar to the histogram
of Model~12.
}
\end{figure}

While the minimal model is the previously published quantitative model,
the poor agreement
was expected by the BATSE Team, e.g., the 
3B catalog [5] notes
``However, the various tests yield only an estimate of the average systematic
error.    A small fraction of the locations could be substantially worse
than the average, i.e., the location error distribution may have a 
non-Gaussian tail.''
We now have enough accurate comparison locations to test models with
additional parameters.

\begin{table}[tb!]
\begin{center}
\caption{The Models}
\begin{tabular}{clclc}
  
Model &    Description    &   Log$_{10}$ &  Parameter  &    Odds \\
      &                &   Likelihood    &   Values  &     Ratios  \\
      &                         &                \\
\tableline
\vspace{2.5mm}
  0   &    Minimal              &                 140.8   &
     $\sigma_{\rm sys} = 1.6^\circ$ fixed  \\
\vspace{2.5mm}
  1   &  One $\sigma_{\rm sys}$  &        215.0   &
     $\sigma_{\rm sys} = 3.07^\circ \pm0.12^\circ$  &
   $O_{1/0} \approx 8 \times 10^{72}$    \\
  4   &  $\sigma_{\rm sys}= A (\sigma_{\rm stat}/1^\circ)^\alpha $  &  217.4 &
        $A=3.22^\circ \pm 0.13^\circ$  &
   $O_{4/1} \approx 25$    \\
\vspace{2.5mm}
  & & &        $\alpha=0.18 \pm 0.05$  &
   $O_{4/0} \approx 2 \times 10^{74}$  \\
  2   &  Core plus Tail        &   226.7               &
        $f_1 = 0.73 \pm 0.08$  &
   $O_{2/4} \approx 6 \times 10^{8}$    \\
      & & & $\sigma_{{\rm sys},1}= 1.9^\circ \pm 0.2^\circ$  &
      $O_{2/1} \approx 2 \times 10^{10}$  \\
\vspace{2.5mm}
      & & & $\sigma_{{\rm sys},2} = 5.4^\circ \pm 0.7^\circ$ &
   $O_{2/0} \approx 1 \times 10^{83}$  \\
 12   &  Datatype   &  229.9  &
         $f^1_{\rm CONT} = 0.76 \pm 0.07$            &
   $O_{12/2} \approx 100$    \\
      & Dependence & &  $\sigma^1_{\rm CONT} = 1.6^\circ \pm 0.2^\circ$   &
   $O_{12/4} \approx 6 \times 10^{10}$   \\
      & & &   $\sigma^2_{\rm CONT} = 5.4^\circ \pm 0.8^\circ$ &
   $O_{12/1} \approx 2 \times 10^{12}$  \\
\vspace{1.5mm}
      & & & $\sigma_{\rm other} = 3.7^\circ  \pm 0.3^\circ $  &
    $O_{12/0} \approx 1 \times 10^{85}$  \\
\end{tabular}
\end{center}
\end{table}

%Model~2 quantifies the idea of a Gaussian distribution with an extended tail
%with the sum of two Fisher distributions with differing systematic errors.

Model~2 uses the sum of two Fisher distributions with differing 
%systematic errors
values of $\sigma_{\rm sys}$
to implement a Gaussian distribution with an extended tail.
It  is a very major improvement over the simpler models (Table~1).
Model~2 is an uncorrelated model---it does not use any datum
to determine the value of $\sigma_{\rm sys}$ for a location and
is best thought of as a Gaussian with wings rather than a
two-component model because BATSE data does not assign a burst to a
component.

So far, the only significant correlation we have found
between location errors and other location properties
is with the datatype used to
locate the event.
Fitting Model~1, one value of $\sigma_{\rm sys}$, to
locations obtained with the 16-channel CONT datatype,
the best fit value of $\sigma_{\rm sys}$ is $2.81^\circ \pm 0.14^\circ$,
while for the other datatypes, the best value is $3.69^\circ \pm 0.29^\circ$,
clearly showing that locations obtained using 16-channel data are
more accurate than those obtained with 4- or 1-channel data.
%Events are located with 4- or 1-channel data when CONT data are unavailable
%due to a telemetry gap or to obtain a higher SNR when the event is shorter
%than the 2~s resolution of the CONT data.

Based upon this correlation, Model~12 was constructed:
events located with CONT data have a core-plus-tail
distribution similar to that of Model~2, while events located using
other datatypes have a single value for $\sigma_{\rm sys}$ (Table~1).\
The data and Model~12 agree well (Fig.~1).
The evidence for an extended tail for the ``other'' locations is 
marginal (odds ratio of 4).
%probably because there are too few
%``other'' locations (120) to demonstrate the existence of a tail.
So far extensions of Model~12, such as adding intensity dependences,
have at best marginal odds ratio improvements ($\leq 10$) and frequently
unconvincing parameter values.

We find only weak evidence for a direct dependence of $\sigma_{\rm sys}$ on the
intensity of a burst, as measured either by $\sigma_{\rm stat}$ or
by fluence.
Model~4, with $\sigma_{\rm sys}$ a power law of $\sigma_{\rm stat}$
[3],
is a definite improvement over Model~1.
However, an uncorrelated model is yet superior, indicating
that the intensity correlation of Model~4 is not supported by the current data.
Additionally, a histogram (not shown) 
shows Model~4 to be a poor fit to the data.
Consequently, the extrapolation
of Model~4 to faint bursts (e.g., $\sigma_{\rm sys} = 8^\circ$ for
$\sigma_{\rm stat} = 10^\circ$) [3], and 
the consequent conclusion that the 3B catalog
has larger systematic errors for faint bursts than the 1B catalog [3]
are not justified.
%Similar results are obtained for their fluence-dependent model,
%which is not included in Table~4 because it can only be applied to the
%smaller sample of bursts with fluence measurements.

While we do not find an intrinsic correlation of $\sigma_{\rm sys}$
with burst intensity, the fraction of events located with CONT data
decreases with decreasing intensity, thereby causing an indirect intensity
correlation via the datatype correlation.
The fraction of CONT locations falls from 70\% for events with
$\sigma_{\rm stat} < 3.7^\circ$ to 40\% for the remaining events.
However, this indirect correlation
does not cause $\sigma_{\rm sys}$ to scale with intensity,
instead it causes
the fraction of events with $\sigma_{\rm sys} = 3.7^\circ$ to be higher
for sets of fainter bursts.

%\section*{Model Implementation}

While the odds ratio favors Model~12 over Model~2,
Model~2 is probably sufficient
for most purposes.   
Here we give a detailed
specification of Model~2 so that others can implement these two models.

Model~2 consists of two terms:
\begin{equation}
P = f_1 P_1 + (1-f_1) P_2
\end{equation}

Each term is a Fisher probability distribution so that the
probability of the true location lying in a ring of $\gamma_1$ to
$\gamma_2$ about the BATSE location is:
\begin{equation}                   
P_i = 
\frac {\kappa_i} {2 \pi ({\rm e}^{\kappa_i} - {\rm e}^{-\kappa_i})}
   \int_{\Omega} d\Omega \: {\rm e}^{\kappa_i \cos \gamma}
 =  \frac {\kappa_i} {{\rm e}^{\kappa_i} - {\rm e}^{-\kappa_i}}
\int_{\gamma_1}^{\gamma_2} d\gamma \sin \gamma \:
               {\rm e}^{\kappa_i \cos \gamma}
\end{equation}

By definition, the radius $\sigma_{{\rm tot},i}$ contains 68\% of the
probability.    Setting $P_i=0.68$, $\gamma_1=0$, and
$\gamma_2= \sigma_{{\rm tot},i}$, we obtain (with
$\sigma_{{\rm tot},i}$ in radians):
\begin{equation}
     \kappa_i = \frac {1} {(0.66 \sigma_{{\rm tot},i})^2 }
\end{equation}

For each burst, the
values of the total location errors $\sigma_{{\rm tot}, i}$ are
calculated from the statistical error $\sigma_{\rm stat}$  listed in the 
catalog and the model errors $\sigma_{{\rm sys},i}$:
\begin{equation}
   \sigma_{{\rm tot},i} =
     [\sigma_{\rm stat}^2 + \sigma_{{\rm sys},i}^2]^{0.5}
\end{equation}

\section*{Conclusions}

The best models are 2 and 12, which are implementations of the
BATSE Team's expectation that the location error distribution is
Gaussian modified with an extended tail.
Model~12 incorporates the
only  significant correlation that we have identified, basing the
error distribution on the datatype used to derive the location.

While the fraction of locations belonging to the larger systematic error
term is 27\% according to Model~2,  this does not mean that
27\% of the locations have large errors---the large systematic error
term has a significant probability of a small location
error.    
For events with $\sigma_{\rm stat} \ll \sigma_{{\rm sys},1}$,
the fraction of locations with an
error past $2 \sigma_{{\rm sys},1} = 3.8^\circ$ is 16\%.

The results herein are upper-limits
on BATSE systematic errors since they are based on the assumption
that there are no errors in the comparison dataset.
As part of this effort, most of the outliers were re-examined and
eight significant  revisions were made to the preliminary IPN catalog.
Even a single outlier past $4\sigma$ can significantly effect
the error model parameter values.

We continue 
%to search for additional correlations and
to test additional models.
Post-4B improvements in the location algorithm
will lead to revised error model parameters.

\end{document}